\documentclass[prb,preprint,showpacs,preprintnumbers,amsmath,amssymb]{revtex4}

\usepackage{graphicx}

\usepackage{dcolumn}

\usepackage{bm}

\newcommand{\gma}{$\rm{Ga_{1-x}Mn_{x}As}$}
\newcommand{\gia}{$\rm{Ga_{1-x}In_{x}As}$}
\newcommand{\rxx}{$R_{\rm{xx}}$}
\newcommand{\rxy}{$R_{\rm{xy}}$}
\newcommand{\tc}{$T_{\rm{C}}$}

\begin{document}
\title{Magnetoresistance Anomalies in (Ga,Mn)As Epilayers with Perpendicular Magnetic Anisotropy }
\author{G.\ Xiang$^1$}
\author{A. W. Holleitner$^2$}
\author{B. L. Sheu$^1$}
\author{F. M. Mendoza$^2$}
\author{O. Maksimov$^1$}
\author{P. Schiffer$^1$}
\author{D. D. Awschalom$^2$}
\author{N.\ Samarth$^{1,2}$}\email{nsamarth@psu.edu}
\affiliation{$^1$ Physics Department $\&$ Materials Research Institute, Penn State University, University Park PA 16802\\
$^2$Center for Spintronics and Quantum Computation, University of California, Santa Barbara CA 93106}

\begin{abstract}
We report the observation of anomalies in the longitudinal magnetoresistance of tensile-strained \gma~ epilayers with perpendicular magnetic anisotropy. Magnetoresistance measurements carried out in the planar geometry (magnetic field parallel to the current density) reveal ``spikes" that are antisymmetric with respect to the direction of the magnetic field. These anomalies always occur during magnetization reversal, as indicated by a simultaneous change in sign of the anomalous Hall effect. The data suggest that the antisymmetric anomalies originate in anomalous Hall effect contributions to the longitudinal resistance when domain walls are located between the voltage probes. This interpretation is reinforced by carrying out angular sweeps of $\vec{H}$, revealing an antisymmetric dependence on the helicity of the field sweep.

\end{abstract}
\maketitle

Contemporary interest in spintronics\cite{wolf_science_2001}
provides a strong motivation for understanding the interplay between electrical transport and domain walls (DWs) in both metallic ferromagnets
\cite{kent_review,ono_science_1999,allwood_science_2002,cheng_PRL_2005} and in ferromagnetic 
semiconductors. \cite{tang_prl_2003,tang_nature_2004,holleitner_apl_2004,yamanouchi_nature} Despite a long history of magnetoresistance (MR) measurements in a variety of ferromagnetic materials, ongoing studies continue to uncover new effects such as an antisymmetric MR in metallic ferromagnets with perpendicular magnetic anisotropy \cite{cheng_PRL_2005} and a giant planar Hall effect in ferromagnetic
semiconductors with in-plane magnetic anisotropy.\cite{tang_prl_2003} The ``canonical" ferromagnetic semiconductor \gma~ is of particular interest in the latter context because its physical properties are the focus of extensive ongoing experimental and theoretical studies.\cite{spinbook,samarth_SSP,nature_mat}

Here, we describe novel anomalies observed in MR
measurements of tensile-strained \gma~ epilayers with a perpendicular magnetic anisotropy. Although the background MR in these samples is symmetric with respect to the direction of the magnetic field $\vec{H}$, we also observe resistance ``spikes" with an antisymmetric deviation $\Delta R$ from the background MR i.e. $\Delta R(H) = -\Delta R(-H)$, similar to recent observations in metallic ferromagnetic multilayers with perpendicular anisotropy.\cite{cheng_PRL_2005} 
The anomalies always occur during magnetization reversal -- indicated by a change in sign of the anomalous Hall effect (AHE) -- and hence suggest an intimate connection with the nucleation of DWs. Unlike the recent observations of antisymmetric MR in metallic multilayers, we observe the field antisymmetric \rxx$(H)$ only in magnetic field sweeps carried out in the planar geometry, where $\vec{H}$~ is nominally applied in the epilayer plane and parallel to the current density ($\vec{j\\\\\\\\}$). Despite this difference, our data strongly suggest that the antisymmetric anomalies have the same origin as in the metallic multilayer studies referenced above: circulating currents near DWs located between the voltage probes produce AHE contributions to \rxx, an effect clearly enhanced during magnetization reversal. The field antisymmetry is then a direct consequence of the intrinsic field-dependence of the AHE. (We note that such circulating currents are also known to create a strong \rxy~ contribution to measurements of \rxx~ in ferromagnetic semiconductors with in-plane anisotropy.\cite{tang_nature_2004,tang_condmat04}) Our interpretation of the data is reinforced by carrying out {\it angular} sweeps of $\vec{H}$ at constant field magnitude, revealing MR spikes that have an antisymmetric dependence on the helicity of the field sweep. This unusual feature has a straightforward explanation within the circulating current picture.

We have fabricated and measured several \gma~ samples with perpendicular magnetic anisotropy. Full details of the sample growth by molecular beam
epitaxy and post-growth annealing procedures are described
elsewhere.\cite{maksimov_jcg_2004} The anomalous magneto-transport
behavior discussed in this paper is generic to all the samples
studied (both as-grown and annealed). We present data from two
annealed samples (A and B) and one as-grown sample
(C) with \tc$= 135 \rm{K}, 130 \rm{K}$ and $125 \rm{K}$,
respectively. All samples are grown on (001) semi-insulating GaAs
substrates after the deposition of a buffer heterostructure that
consists of 100 nm GaAs (grown at $630^{\circ}$C), followed by 1$\mu$m
strain-relaxed  \gia~ (grown at $480^{\circ}$C). The magnetically active
layer consists of a 30 nm thick \gma~ epilayer grown  on the \gia~
surface at $230^{\circ}$C. This creates a coherent in-plane tensile strain in the \gma, resulting in perpendicular magnetic anisotropy, with the easy axis along [001]. After removal from the MBE chamber, the wafers
are cleaved into smaller pieces and some of these pieces are
annealed for 2 hours at $250^{\circ}$C. As found for standard \gma~ grown
on GaAs,\cite{hayashi_APL_2001,potashnik_APL_2001} the low
temperature annealing increases the Curie temperature in these
tensile strained samples. Atomic force microscopy shows that the
grown surface has a cross-hatched pattern with ridges running along
the $[ 1 \bar{1} 0 ]$~ direction that are higher than those running
along the $[110]$~ direction; the lateral spacing between these
ridges is $\sim 1 \mu$m.

We measure the longitudinal and Hall resistance (\rxx~ and \rxy, respectively) in mesa-etched $800 \mu$m $\times 400 \mu$m
Hall bars patterned using conventional photolithography and a
chemical wet etch, with electrical contacts made using indium solder
and gold wire leads. For convenience, each Hall bar pattern
consists of three arms oriented along three principal
crystalline directions $[110]$, $[010]$~ and $[1 \bar{1} 0]$, as shown in Fig. 1. Both
dc as well as low-frequency ac magneto-transport measurements are
carried out from 300 mK to 300 K in different cryostats with
superconducting magnets, including one that contains a vector magnet
that allows angular variations of a 10 kOe field over the unit sphere. We
also measure magnetization ($M$) as a function of temperature and magnetic field in a Quantum Design superconducting quantum interference device (SQUID) magnetometer.

We first discuss the temperature-dependence of \rxx~ and \rxy~ at $H = 0$. This is shown in Fig. 2(a) for sample A with the current density $\vec{j} ||[ 1 \bar{1} 0
]$.  Temperature- and
field-dependent SQUID measurements on a separate piece of this sample indicate the onset of ferromagnetism at \tc$ = 135$ K with an easy axis along [001]. The zero-field behavior of \rxx~ and \rxy~ confirms the SQUID measurements: \rxx~ has a maximum in the
vicinity of \tc, while \rxy~
reveals the emergence of the AHE below \tc.\cite{spinbook} A detailed analysis of the temperature dependence of \rxy~ and $M$ at zero field yields $R_{\rm{xy}} \propto
R_{\rm{xx}} M$ over the temperature range $20 \rm{K} \leq T \leq 120 \rm{K}$, empirically consistent with AHE in the presence of skew-scattering.\cite{hurd_book_1972,smit_physica_1955} At lower temperatures, the temperature-dependence does not fit any known pictures of the AHE.   

We now discuss the magnetic field dependence of \rxx~ and \rxy~ for the same sample in the perpendicular geometry, where the magnetic field is perpendicular to both the current density and the sample plane (see Fig. 1(a)). The lower traces in Figs. 2(b) and (c) show \rxy$(H)$ and \rxx$(H)$, respectively, at $T = 90$K with $\vec{H} || [001]$ and $\vec{j} || [1 \bar{1} 0]$). The field-dependence of \rxy~ is
dominated by the AHE and shows an
easy axis hysteresis loop that is proportional to $M(H)$, yielding a coercive field identical to that obtained in SQUID measurements ($H_{\rm CE} \sim 20$
Oe) . While the field-dependence of \rxx~ shows reproducible features (hysteresis and discontinuous resistance jumps) associated with magnetization reversal, these effects are relatively small and difficult to resolve, particularly at high temperatures.

In contrast to the data measured in the perpendicular geometry, we find surprising characteristics in the magnetic field dependence of both \rxy~
and \rxx~ in the planar geometry, where the magnetic field is in the sample plane and parallel to the current density (see Fig. 1(a)). The upper traces in Figs. 2(b) and 2(c) show \rxy$(H)$ and \rxx$(H)$, respectively, with $\vec{j} || \vec{H} || [1 \bar{1} 0]$.\cite{note2}  The hysteretic Hall resistance loop now has an unusual shape and changes sign at a field
$H_{\rm{CH}} \sim 2$ kOe -- almost two orders of magnitude larger than
the coercive field $H_{\rm{CE}}$~ observed in the easy axis configuration (Fig. 2(b)). Moreover, we find striking resistance ``spikes" in \rxx~ that are {\it
antisymmetric} in magnetic field direction and that occur when
when \rxy$= 0$. This antisymmetry is at first sight surprising because it appears to run counter to the Onsager reciprocity requirement that \rxx$(H) =$\rxx$(-H)$.  Although asymmetric MR has been the focus of many discussions within the context of mesoscopic transport,\cite{buttiker_PRL_1986,benoit_PRL_1986} macroscopic samples are not expected to exhibit such asymmetric behavior. Figure 3 presents
additional field dependent measurements of \rxy and \rxx at
different temperatures showing that the characteristics observed in
Fig. 2(c) are generic and can be readily observed at temperatures as high as 100 K. 

The unusual shape of the hysteretic Hall loops observed in the planar geometry can be explained by assuming that the nominally in-plane magnetic field is slightly misaligned, resulting in an AHE contribution from the magnetization component along the easy axis ($\hat{z}$). To illustrate the physics underlying the Hall loop, we follow \rxy$(H)$ from negative to positive magnetic field, starting at $H = -10$kOe nominally applied in-plane and parallel to $\vec{j}$, but with a small unintentional misalignment towards $+\hat{z}$ (characterized by an angle $\delta$~ between $\vec{H}$ and $\vec{j}$.  At this large magnetic field, $M_z << M_{\rm{in-plane}}$, resulting in a negligible AHE, consistent with the data in Figs. 2(b) and 3(a).
As $|\vec{H}| \rightarrow 0$, $\vec{M}$
gradually changes orientation from in-plane
($-\hat{x}$) towards $+\hat{z}$ (the easy axis), with the symmetry
between $+\hat{z}$~ and $-\hat{z}$~ broken by the field misalignment. This results in a
positive value of \rxy~ at $H = 0$.  When we reverse the direction of $\vec{H}$ and increase its magnitude,
the misalignment of $\vec{H}$~ is now towards $-\hat{z}$. This
initiates a reversal of the sample magnetization via the nucleation and propagation of DWs, and is accompanied by a change in sign of \rxy. While we do not presently have any direct information about the domain structure in these samples, scanning Hall probe imaging of similar samples has shown the presence of striped domains with widths that are as large as $\sim30 \mu \rm{m}$.\cite{shono_APL_2000} Further
increases in the magnitude of $\vec{H}$~ gradually orient the magnetization in-plane, so that the AHE decreases and eventually becomes negligible. We note that while we might expect additional contributions to \rxy~ from the planar Hall effect, these appear to be small compared with the AHE in samples with tensile strain.

The Hall loop provides a crucial insight into the different behavior of \rxx$(H)$ in the planar and perpendicular geometries:  the data show that large-scale switching occurs in the planar geometry only when 
$H_z \geq {H_{\rm{CH}}} \sin \delta = H_{\rm{CE}}$.  Using the coercive fields observed for the perpendicular and planar geometries (Fig. 2(b)), we estimate that the misalignment angle $\delta \sim 0.6^{\circ}$. Since the typical magnetic field sweep rate is 10 Oe/s, the z-component only increases at a rate of $\sim  0.1$ Oe/s. Consequently, magnetization reversal in the planar geometry case occurs quasi-statically, allowing the slow propagation of DWs during the magnetic field sweep. This is in strong contrast with the perpendicular geometry case where magnetization reversal occurs very rapidly (in a couple of seconds) with our typical sweep rates.
The antisymmetric jumps in \rxx~ in the planar geometry can now be qualitatively understood using the picture developed in Ref. 5. When the magnetization is homogeneous, \rxx$(H) =$\rxx$(-H)$: this situation applies to most of the magnetic field regime in Figs. 2 (c) and 3(b). However, when DWs are formed, creating a spatially inhomogeneous magnetization, the current density is perturbed by (static) circulating currents that flow in the vicinity of DWs.\cite{cheng_PRL_2005,tang_condmat04} These circulating currents lead to an admixture of \rxy~ into measurements of \rxx. Because \rxy$(H) = -$\rxy$(-H)$, the contribution is antisymmetric with respect to the field direction. To test the correctness of this interpretation, we have carried out simultaneous measurements of \rxx~ using contacts on opposite sides of a Hall bar which should show opposite signs for the circulating current contribution to \rxx. Our measurements indeed reveal this asymmetry, showing that the sum ${R_{xx}}^{\rm{up}}(H) +{R_{xx}}^{\rm{down}}(H)$ is symmetric with respect to field reversal. Further,  we find that the data follow the ``sum rule" ${R_{xx}}^{\rm{up}} - {R_{xx}}^{\rm{down}} = {R_{xy}}^{\rm{left}} - {R_{xy}}^{\rm{right}}$,\cite{tang_nature_2004} again consistent with the picture that the resistance anomalies arise from transverse electric field contributions due to circulating currents. 

Further evidence for AHE contributions to \rxx~ emerges from MR measurements carried out as a function of the magnetic field angle.   Figure 4(a) shows angle-dependent measurements of \rxx~ in a different sample (B) when the magnitude of $\vec{H}$~ is held constant
while its direction is changed using a vector magnet (see for example Fig. 1(b)). The data are shown for Hall bars patterned along two different crystalline directions ($\vec{j}
|| [110]$~ and $\vec{j} || [1\bar{1}0]$), while we sweep the azimuthal angle $\theta$ through $2 \pi$
in the plane normal to the current density. As in the field-swept data, the angular sweeps also show spikes in \rxx~ whenever \rxy~ (and hence $\vec{M}$) changes sign (Fig. 4(b)). The spikes have an asymmetric dependence on the helicity of the sweep. While this is initially surprising, the reason becomes clear if we assume that the field rotation nucleates DWs just as in the case of a field sweep: when $\vec{H}$ is swept in the $yz$-plane, DWs are nucleated whenever $H_z > H_{\rm{CE}}$. For sweeps of opposite helicity, the magnetization reversal at $-\theta$ and $+\theta$ always occur towards the same direction, hence creating an AHE contribution of the same sign. 
 
The angular scans also reveal that the DW structure during these angular sweeps is quite complex. Figure 4(c) shows high-resolution angle-dependent 
magneto-resistance measurements at 4.2 K with angular steps of $0.1^\circ$ on another sample (C) patterned along $[0 \bar{1} 0]$.  The reproducible, fine structure in each of
the resistance spikes, suggests the presence of small magnetic domains. It is pertinent to question whether the resistance spikes indicate a transient state or a stable one. To examine this issue, we carried out time-dependent measurements of \rxx~ after sweeping the magnetic field angle to a configuration where
a resistance spike approximately reaches its maximum value and then holding the field direction and magnitude fixed for several minutes.  Our measurements show that the value of
\rxx~ does not change over this time scale, indicating that the magnetic state
creating the resistance spike is a stable one. These observations also confirm that
the resistance spikes do not have an extrinsic origin such as inductive
effects.

In summary, we have observed field-antiymmetric anomalies in the longitudinal MR of \gma~ epilayers with perpendicular magnetic anisotropy. These antisymmetric anomalies arise from AHE contributions to \rxx~ when DWs are located in between the measuring contacts, but are only observed in the planar geometry when the magnetic field is parallel to the current density. This contrasts with similar observations in metallic ferromagnetic multilayers with perpendicular magnetic anisotropy where such antisymmetric MR is observed with the field normal to the current and the sample plane. Our observations are likely to be important for the analysis of MR measurements in experiments focusing on the control and measurements of DWs in laterally patterned structures fabricated from tensile-strained \gma.

This research has been supported by grant numbers ONR N0014-05-1-0107, DARPA/ONR N00014-99-1-1093, -99-1-1096, and -00-1-0951, University of California-Santa Barbara subcontract KK4131, and NSF DMR-0305238, -0305223 and -0401486. 

\begin{center}
{\bf References}
\end{center}

\begin{thebibliography}{22}
\bibitem{note2} Because
the magnetic field is applied along a hard axis in this geometry,
cooling down a sample in zero applied field and then starting
magnetic field sweeps can create unstable magnetic domain patterns.
Hence, to produce consistently reproducible data when the magnetic
field is then swept through a loop, we carry out magnetoresistance
measurements after cooling the sample down from room temperature in
an in-plane magnetic field of $- 20$ kOe.
\end{thebibliography}

\newpage

\begin{center}
{\bf Figure Captions}
\end{center}

Fig. 1 (a) Schematic depiction of Hall bar orientation. The field directions $\vec{H}_{\perp}$ and $\vec{H}_{||}$ relative to the current density $\vec{j}$ correspond to the perpendicular and planar geometries discussed in the manuscript. (b) Schematic depiction of one of the angular field sweeps discussed in the text.

Fig. 2 (a) Temperature dependence of \rxx~ and \rxy~ for sample A with $\vec{j} || [1\bar{1}0]$ in the absence of an external magnetic field. Panels (b) and (c) show the magnetic field dependence of \rxy~ and \rxx, respectively, at $T = 90$K for the same sample in the planar geometry ($\vec{H} || [1\bar{1}0]$, upper trace) and the perpendicular geometry ($\vec{H} || [001]$, lower trace). In both cases, the Hall bar is oriented such that $\vec{j} || [1\bar{1}0]$. The data in (b) and (c) are offset along the y-axis for clarity. 

Fig. 3. Magnetic field dependence of (a) \rxy~ and (b) \rxx~ in sample A at $T = 40, 60, 80, 100$K. The data are all taken in the planar geometry with $\vec{H} || \vec{j} || [1\bar{1}0]$. The data are offset along the y-axis for clarity.

Fig. 4. Panels (a) and (b) show \rxx~ and \rxy, respectively, $T = 90$K for sample B as $\vec{H}$ is swept through an angle $\theta = 2 \pi$ in the plane perpendicular to the current density with $|\vec{H}| = 730$G. The angle $\theta$ is defined between $\vec{H}$ and [001]. The upper traces correspond to $\vec{j} || [110]$, while in the lower traces $\vec{j} || [1\bar{1}0]$.  The data are offset along the y-axis for clarity. Panel (c) shows measurements of \rxx~ taken with higher angular resolution on another sample (C) at $T = 4.2 K$. Here, $\vec{j} || [010]$ and $\vec{H}$ is again swept through $\theta = 2 \pi$ in the plane perpendicular to the current with $|\vec{H}| = 730$G.

\newpage
\begin{figure}[h]
\begin{center}
\includegraphics[]{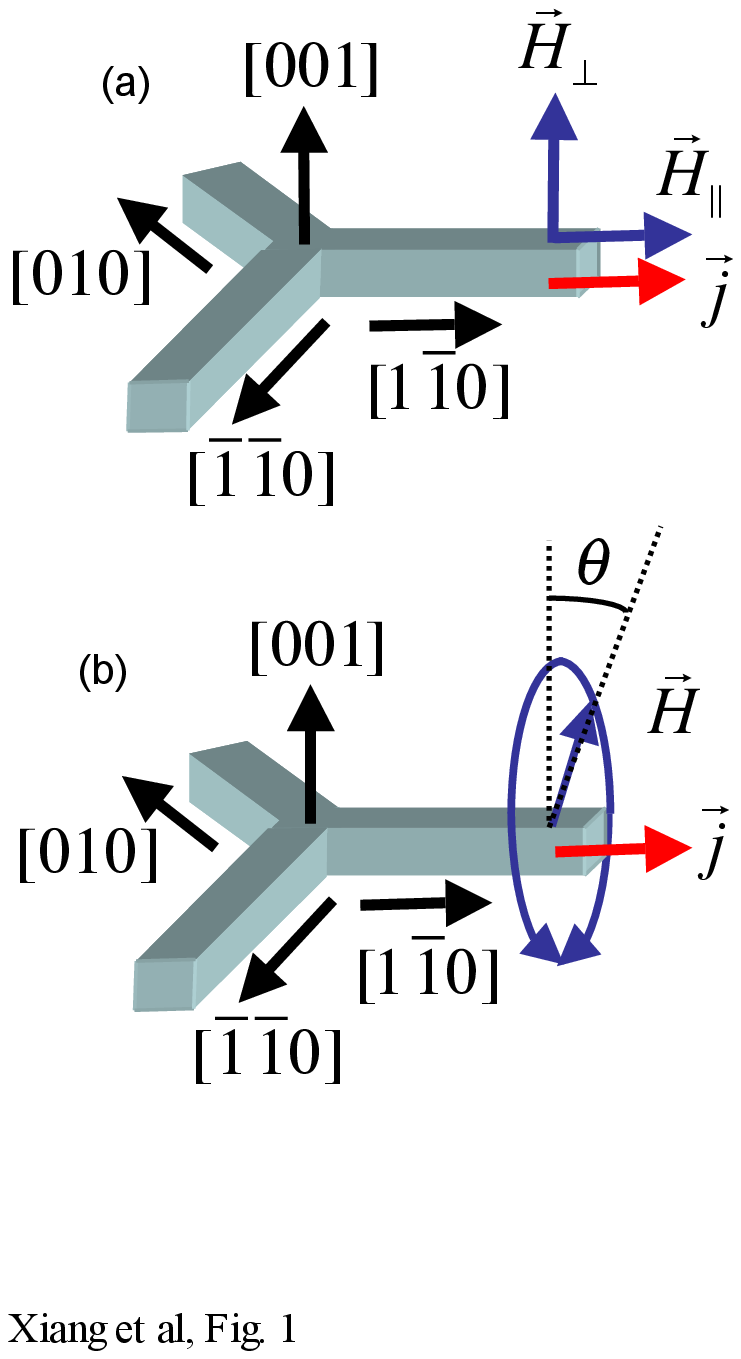}
\label{gang_fig1}
\end{center}
\end{figure}

\newpage
\begin{figure}[h]
\begin{center}
\includegraphics[]{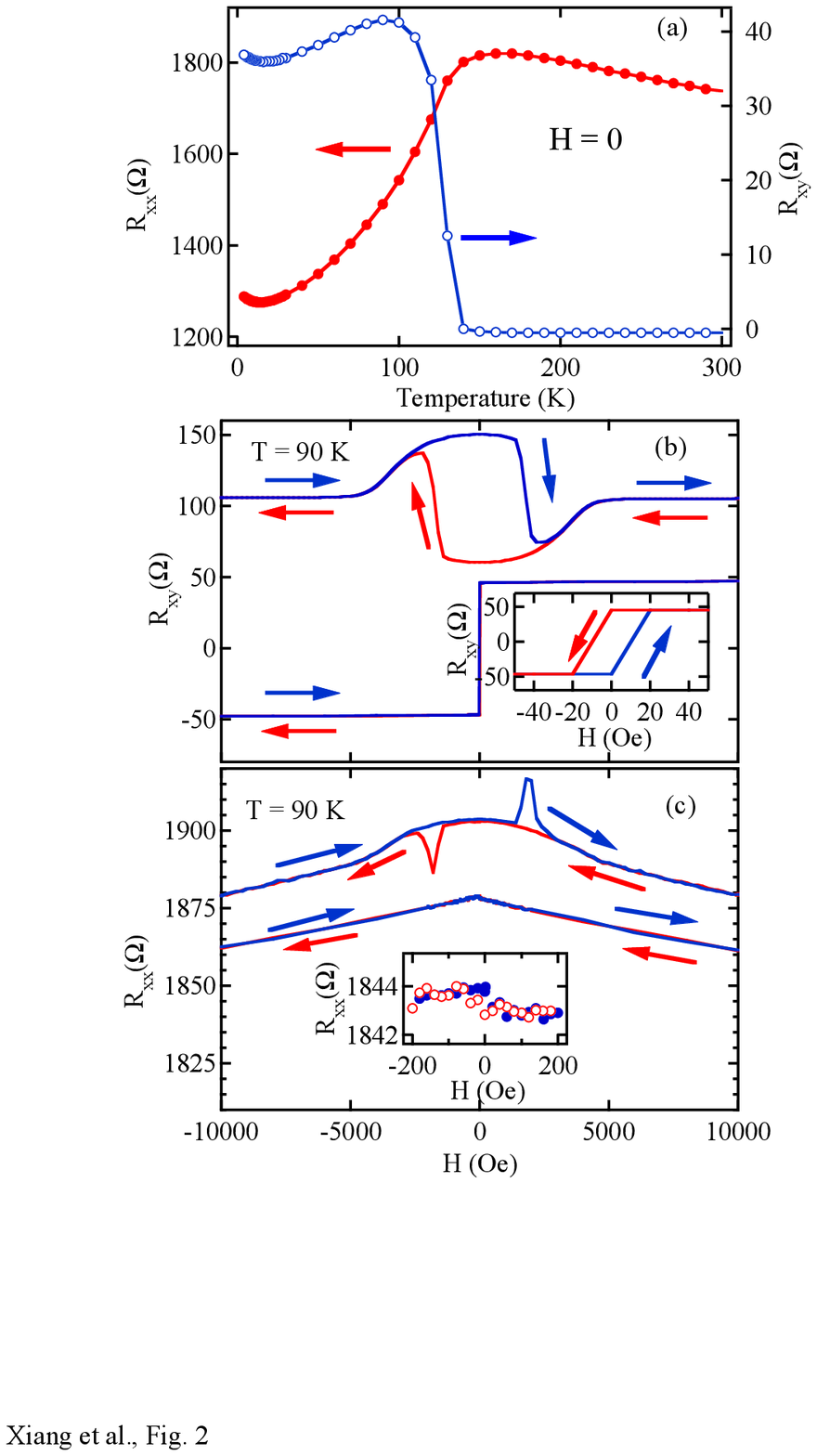}
\label{gang_fig2}
\end{center}
\end{figure}

\newpage
\begin{figure}[h]
\begin{center}
\includegraphics[]{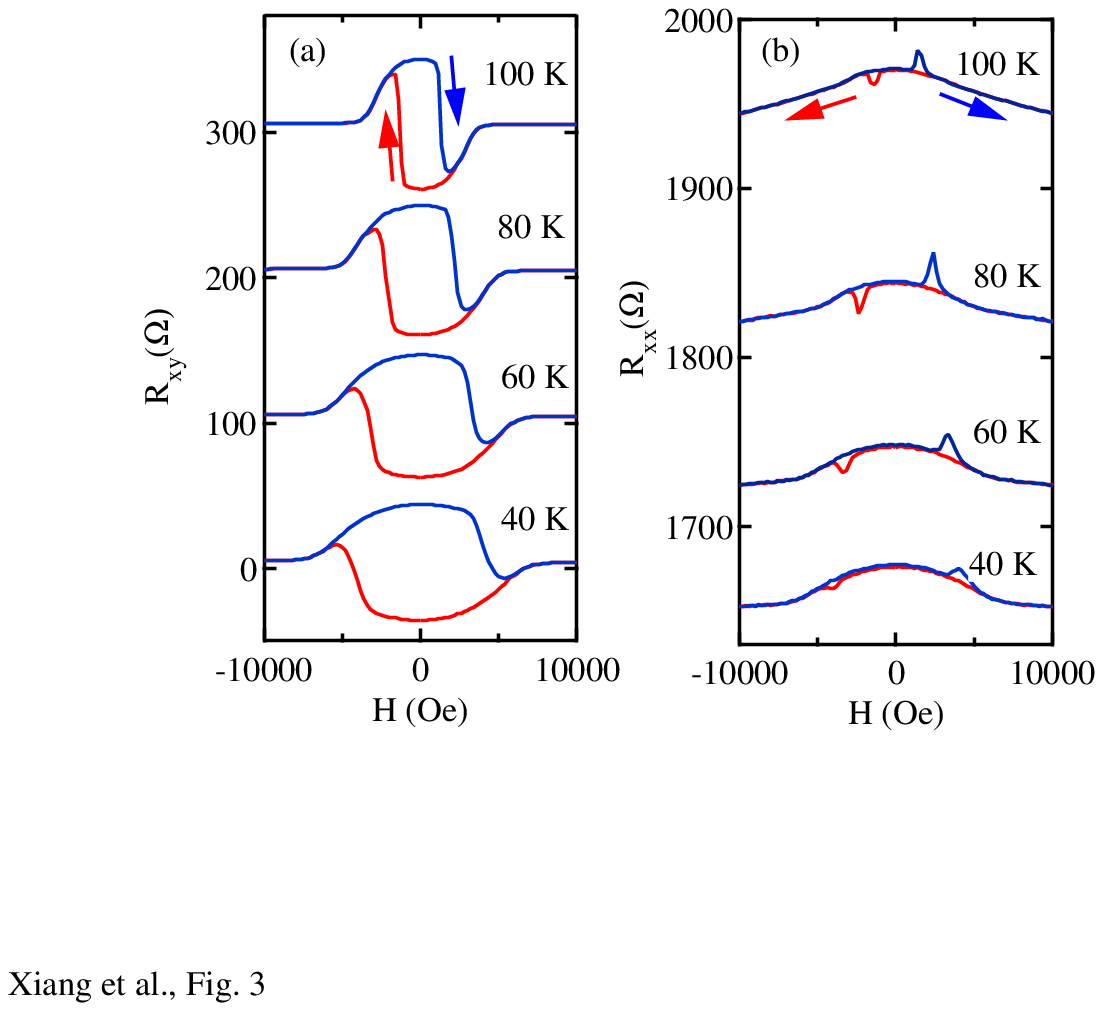}
\label{gang_fig3}
\end{center}
\end{figure}

\newpage
\begin{figure}[h]
\begin{center}
\includegraphics[]{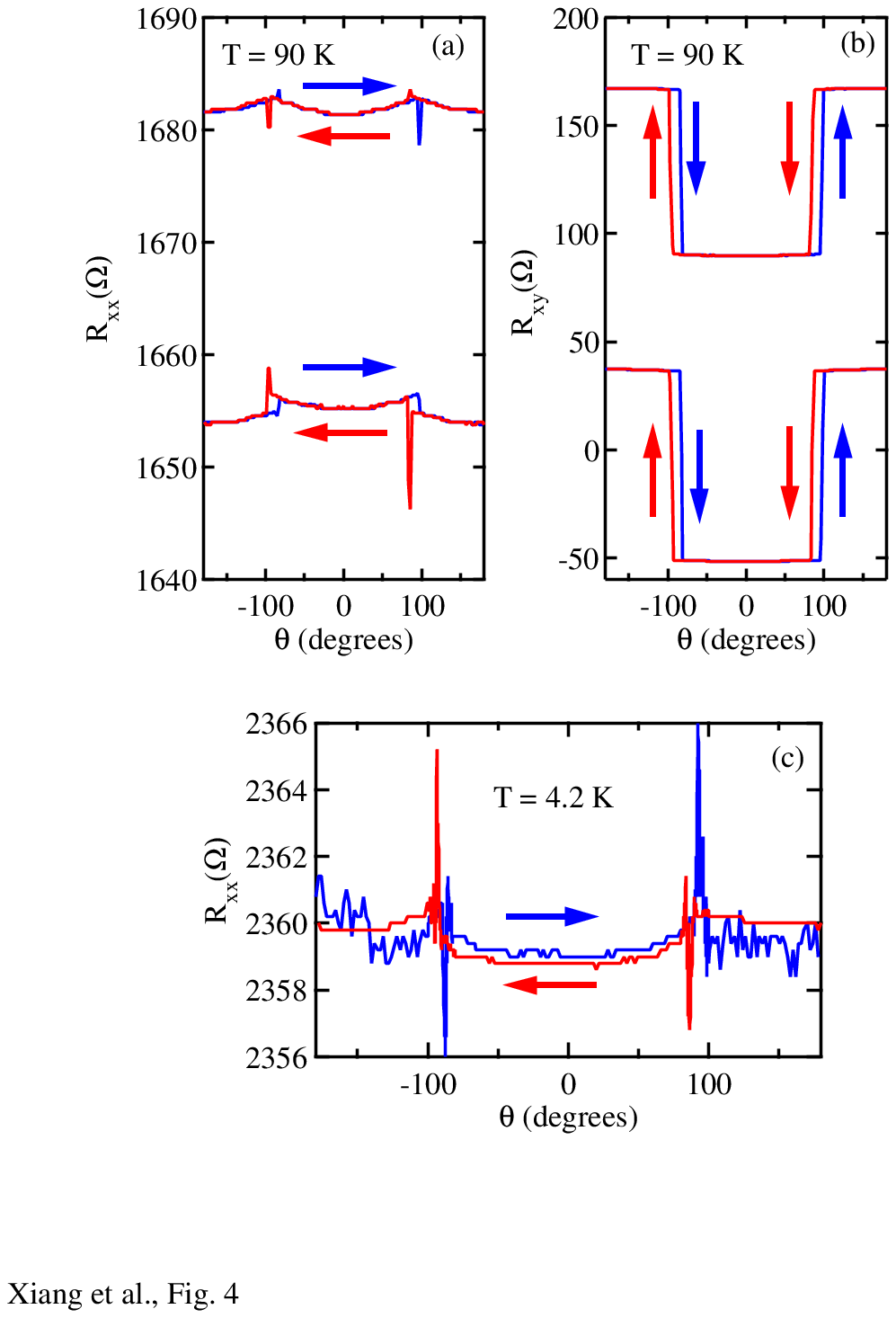}
\label{gang_fig4}
\end{center}
\end{figure}


\end{document}